\documentclass[preprint,showpacs,preprintnumbers,amsmath,amssymb]{revtex4}
\usepackage{graphicx}
\usepackage{dcolumn}
\usepackage{bm}
\usepackage{hyperref}

\begin{document}
\preprint{}
\title{Conformal Gravity from AdS/CFT mechanism}
\author{Rodrigo Aros}
\affiliation{Departamento de Ciencias F\'{\i}sicas,\\ Universidad Andr\'es Bello, Av. Rep\'ublica
252, Santiago,Chile}
\author{Mauricio Romo}
\affiliation{Departamento de F\'{\i}sica, Facultad de Ciencias F\'{\i}sicas y Matem\'{a}ticas, Universidad
de Chile, Av. Blanco Encalada 2008, Santiago,Chile}
\author{Nelson Zamorano}
\affiliation{Departamento de F\'{\i}sica, Facultad de Ciencias F\'{\i}sicas y Matem\'{a}ticas, Universidad
de Chile, Av. Blanco Encalada 2008, Santiago,Chile}
\date{\today}

\begin{abstract}
We explicitly calculate the induced gravity theory at the boundary of an asymptotically
Anti-de Sitter five dimensional Einstein gravity. We also display the action that encodes the
dynamics of radial diffeomorphisms. It is found that the induced theory is a four dimensional
conformal gravity plus a scalar field. This calculation confirms some previous results found
by a different approach.
\end{abstract}
\maketitle

\section{Introduction}

A connection between string theory on $AdS_{5}\times S^{5}$ and super Yang Mills theory in
four dimensions was proposed by J. Maldacena some years ago \cite{Maldacena:1998re}. More
recently, this result gave rise to what is currently called the AdS/CFT conjecture. However
the name has been enlarged to include many different results. The AdS/CFT conjecture relates
the renormalized gravity action induced in the boundary with the expectation value of the
stress tensor of the dual CFT as:

\begin{equation}
\frac{1}{\sqrt{\gamma}}\frac{\delta S_{ren}}{\delta\gamma_{ij}}=\langle T_{ij}\rangle_{CFT},
\end{equation}
where $\gamma_{ij}$ is the metric induced on the boundary

As stated today the AdS/CFT conjecture actually represents a realization of holography as
proposed 10 years ago by Susskind and t'Hooft  \cite{Susskind:95}, \cite{'t Hooft:93}. This
conjecture has been extensively checked, in part, because the conformal symmetry is strong
enough to determine many generic results in a CFT without knowing the details of the
particular theory. For instance, one can demonstrate that the thermodynamics of a black hole
in an asymptotically (locally) AdS space reproduces the thermodynamics of a CFT. To our
knowledge this is true for all the theories of gravity with a single negative cosmological
constant (see for instance \cite{Aros:2000ij}). The main reason is that the thermodynamics of
any CFT is almost completely determined by the conformal symmetry.

Furthermore, one can prove that, under certain particular conditions, a gravitational theory
can be induced in a lower dimensional surface at the bulk. The brane-worlds proposed in
\cite{Randall:1999ee} are a realization of these ideas. In \cite{Carlip:2005tz}, using the
same underlying idea, is shown that the Liouville theory arises as the effective theory at
the AdS asymptotic boundary in 2+1 AdS-gravity. If the AdS/CFT conjecture is to be understood
as a duality relation then a classical solution in the bulk should rise to a quantum
corrected solution at the boundary. This result was actually confirmed between $3+1/2+1$
dimensions in \cite{Emparan:2002px}.

An asymptotically (locally) AdS space needs to be treated carefully otherwise one is usually
led to a divergent behavior in the Lagrangian, the conserved charges or/and the variations of
the Lagrangian. Therefore to confirm many of the results of the AdS/CFT conjecture is
necessary to use some (classical) regularization processes together with a proper set of
boundary conditions. The regularization of the conserved charges has been an interesting
field by itself where many relevant results has been found (See for instance
\cite{Aros:1999id,Henneaux:1999ct,Barnich:2001jy}). A generic method to deal with the
divergent behaviors of the actions appears in \cite{Skenderis:2002wp}, where the conjecture
is used to build a method to compute anomalies of CFT's. In this work part of these results
will be used. In particular in five dimensions a finite version of the Einstein-Hilbert
action \cite{Balasubramanian:1999re} reads:

\begin{eqnarray}
I_{grav} &=& \frac{1}{16\pi G}\int_{M} d^{4}xd\rho\sqrt{g}(R-2\Lambda)-\frac{1}{8\pi
G}\int_{\partial M} d^{4}x\sqrt{\gamma}K \nonumber\\
         &-&\frac{3}{8\pi G}\int_{\partial M} d^{4}x\sqrt{\gamma}-\frac{1}{16\pi G}\int_{\partial M}
d^{4}x\sqrt{\gamma}R[\gamma].\label{5dAction}
\end{eqnarray}

We would like to point out that although in this action every term is divergent, the addition
of all of them becomes finite and well behaved. A discussion of a generalization of this
action can be found in \cite{Olea:2006vd}.

In this work we prove that an effective conformal gravity theory arises at the boundary when
the action displayed in equation (\ref{5dAction}) is used as the bulk's theory. We have
extended to five dimensions the work previously done by Carlip \cite{Carlip:2005tz} in three
dimensions.

The theory obtained coincides with the bosonic part of the super conformal gravity that
appears in \cite{Balasubramanian:2000pq} and in \cite{Liu:1998bu}, however the method
employed to reach this result is different. It is worth to stress that there is another
approach, independent of the two already mentioned, to obtain the same result
\cite{Elizalde:1994nz}.

\section{The four dimensional conformal action and the anomaly}

The purpose of this work is to rewrite the action that appears in equation (\ref{5dAction})
and to show that it can be understood as a four dimensional theory under diffeomorphisms that
preserve the asymptotically AdS scaling of the metric. To fulfill this program, we begin with
a general five dimensional asymptotic AdS metric with a Fefferman-Graham-type expansion near
infinity. This yields the following line element:
\begin{equation}
ds^2 = l^{2}d\rho^{2}+ g_{ij}(x,\rho)dx^{i}dx^{j},\label{metric}
\end{equation}
where $\rho \,\rightarrow\,\infty$, defines the asymptotical (locally) AdS region.

The metric $g_{ij}(x,\rho)$ admits the expansion:
\[
g_{ij}(x,\rho)=e^{2\rho}g^{(0)}_{ij}(x)+g^{(2)}_{ij}(x)+e^{-2\rho}g^{(4)}_{ij}(x)-2e^{-2\rho}\rho
h_{ij}(x)+\ldots
\]

Next, we set $l=1$, thus $\Lambda=-6$. With this expansion the Einstein equations can be
solved iteratively. This yields (see reference \cite{Skenderis:2002wp}):
\begin{equation}
Tr(g^{(4)})=\frac{Tr(g^{(2)2})}{4},\textrm{ }\quad
g^{(2)}_{ij}=-\frac{1}{2}(R^{(0)}_{ij}-\frac{1}{6}R^{(0)}g^{(0)}_{ij}),\quad Tr(h)=0,
\end{equation}
where traces are obtained using the metric $g^{(0)}_{ij}$.

The following step consider a coordinate transformation that must leave invariant the
asymptotic form of the metric (\ref{metric}). Using the prescription displayed in
\cite{Carlip:2005tz}, the transformation reads:
\[
\rho\rightarrow\rho+\frac{1}{2}\varphi(x)+ e^{-2\rho} f^{(2)}(x)+\ldots\,,
\]
\begin{equation}\label{NewCoordinates}
x^{i}\rightarrow x^{i}+e^{-2\rho}h^{(2)i}(x)+\ldots\,\, .
\end{equation}
Note that in this new coordinate system $\rho$ and the variable $x_i$ appear factorized.

The boundary, ($\bar{\rho}\, \rightarrow \,\infty)$, is defined as:
\begin{equation}\label{F(X)}
\rho=\bar{\rho}+\frac{1}{2}\varphi(x)+O(e^{-n\bar{\rho}})=F(x),
\end{equation}
therefore the induced metric at the boundary and the unit normal respectively read:
\begin{equation}
\gamma_{ij}=g_{ij}+\partial_{i}F\partial_{j}F\, ,
\end{equation}
\begin{equation}
n^{a}=\frac{1}{\sqrt{1+g^{ij}\partial_{i}F\partial_{j}F}}(-1,g^{ij}\partial_{j}F).
\end{equation}

In this new system of coordinates (\ref{NewCoordinates}) one can write an expansion in powers
of $\rho$ for the determinant $\sqrt{\gamma}$, the extrinsic curvature and the Ricci scalar
near the boundary. The expansions for each one of these geometrical objets are:
\begin{eqnarray} \sqrt{\gamma}&=&
e^{4\rho}\sqrt{g^{(0)}}+\frac{1}{2}\sqrt{g^{(0)}}e^{2\rho}(Tr(g^{(2)})+g^{(0)ij}\partial_{i}F\partial_{j}F)+\frac{1}{2}\sqrt{g^{(0)}}
\left(Tr(g^{(4)})+\frac{1}{4}Tr(g^{(2)})^{2}\right. \nonumber\\
&-&\frac{1}{2}Tr(g^{(2)2})-\frac{1}{4}(g^{(0)ij}\partial_{i}F\partial_{j}F)^{2}+\frac{1}{2}Tr(g^{(2)})g^{(0)ij}\partial_{i}F\partial_{j}F\nonumber\\
&-&\left. g^{(0)ai}g^{(0)bj}g^{(2)}_{ij}\partial_{a}F\partial_{b}F \right)+\ldots,
\end{eqnarray}
\begin{eqnarray}
K &=& \frac{1}{2}Tr(\gamma\pounds_{n}g_{\|})=-4+e^{-2\rho}(g^{(0)ij}\partial_{i}F\partial_{j}F+g^{(0)ij}\nabla^{(0)}_{i}\nabla^{(0)}_{j}F+Tr(g^{(2)}))+e^{-4\rho}(2Tr(g^{(4)}) \nonumber\\
  &-&Tr(g^{(2)2})-\frac{1}{2}Tr(g^{(2)})g^{(0)ij}\nabla^{(0)}_{i}\nabla^{(0)}_{j}F-\frac{1}{2}Tr(g^{(2)})g^{(0)ij}\partial_{i}F\partial_{j}F)
  +\ldots \,\, \mbox{and}
\end{eqnarray}
\begin{eqnarray}
R[\gamma]&=& e^{-2\rho}(R^{(0)}-6g^{(0)ij}\nabla^{(0)}_{i}\nabla^{(0)}_{j}F-6g^{(0)ij}\partial_{i}F\partial_{j}F)+e^{-4\rho}(-g^{(2)ij}R^{(0)}_{ij}-R^{(0)ij}\partial_{i}F\partial_{j}F \nonumber\\
&+&2g^{(2)ij}\nabla^{(0)}_{i}\nabla^{(0)}_{j}F+Tr(g^{(2)})g^{(0)ij}\nabla^{(0)}_{i}\nabla^{(0)}_{j}F-2g^{(2)ij}\partial_{i}F\partial_{j}F \nonumber \\
&+&2Tr(g^{(2)})g^{(0)ij}\partial_{i}F\partial_{j}F)+\ldots \,.
\end{eqnarray}
The Ricci scalar is defined up to total derivatives of the order of $O(e^{-4\rho})$. All
indices are raised and lowered with $g^{(0)_{i\,j}}$.

Here we have defined
\begin{equation}
Tr(\gamma\pounds_{n}g_{\|})=
\gamma^{ij}\partial_{i}x^{\mu}\partial_{j}x^{\nu}(\pounds_{n}g)_{\mu\nu},
\end{equation}
with $\mu=0\ldots 4$ and $x^{4}=\rho$, the derivative  of the coordinates $x^{\mu}$ are given
by:
\begin{equation}
\partial_{i}x^{\mu}=\delta_{i}^{\mu}+\partial_{i}F\delta_{4}^{\mu}.
\end{equation}

We want to use the expansions just described to extract the finite part of the action
(\ref{5dAction}). First we integrate $\rho$ in the five dimensional action (\ref{5dAction}):\\
\begin{eqnarray}
\int_{M} d^{4}xd\rho\sqrt{g}(R-2\Lambda)_{on-shell}&=& -8\int_{\partial M}
d^{4}x\int^{\rho=F} d\rho\sqrt{g}=-8\int_{\partial M} d^{4}x\int^{\rho=F}d\rho(
e^{4\rho}\sqrt{g^{(0)}} \nonumber\\
+\frac{1}{2}\sqrt{g^{(0)}}e^{2\rho}(Tr(g^{(2)}))&+&\frac{1}{2}\sqrt{g^{(0)}}(Tr(g^{(4)})+\frac{1}{4}Tr(g^{(2)})^{2}-\frac{1}{2}Tr(g^{(2)2}))+\ldots)
\nonumber\\
=\int_{\partial M}
d^{4}x(-2e^{4F}\sqrt{g^{(0)}}&+&\frac{1}{3}\sqrt{g^{(0)}}e^{2F}(R^{(0)})-\frac{1}{4}\sqrt{g^{(0)}}(R^{(0)ij}R^{(0)}_{ij}
\nonumber \\
-\frac{1}{3}R^{(0)2})F+\ldots) & &
\end{eqnarray}
The term proportional to $F$ has a divergent term that must be eliminated adding a
counterterm. This regularization procedure has been introduced by Skenderis
\cite{Skenderis:2002wp}.

Finally, evaluating the action on-shell we get:

\begin{eqnarray}
I_{grav}&=&\frac{1}{16\pi G}\int_{\partial
M}\sqrt{g^{(0)}}(-\frac{1}{16}(R^{(0)ij}R^{(0)}_{ij}-\frac{1}{3}R^{(0)2})+\frac{1}{64}(\partial_{i}\varphi\partial^{i}\varphi)^{2} \nonumber\\
&+&\frac{1}{16}\partial_{i}\varphi\partial^{i}\varphi\nabla^{(0)}_{j}\nabla^{(0)j}\varphi-\frac{1}{8}(R^{(0)ij}R^{(0)}_{ij}-\frac{1}{3}R^{(0)2})\varphi+\frac{1}{8}G^{(0)ij}\partial_{i}\varphi\partial_{j}\varphi)
\end{eqnarray}

This expression can be recognized as the action for a 4-dimensional conformal gravity plus an
anomalous part.

It is worth to stress that this result has been obtained before by at least two different
approaches. For instance, Riegert \cite{Riegert:1984kt} arrived to the same expression
considering an action that could take into account and cancel the anomalous terms. Different
approaches converging to the same conclusion give a solid confidence to the result obtained.

\section{Conclusions}

In this work we have proven that a four dimensional conformal gravity can be obtained through
the AdS/CFT mechanism from five dimensional Einstein gravity. We have demonstrated this
explicitly using the Fefferman-Graham expansion and regularizing the action. As expected the
radial diffeomorphisms induces a Weyl transformation on the boundary metric which in turn
produces the anomalous part as is demonstrated in \cite{Manvelyan:2001pv}. The degrees of
freedom associated to radial diffeomorphisms are encoded in the dynamics of the scalar field
$\varphi$. This action was obtained by Riegert \cite{Riegert:1984kt} as the local form of the
action which gives a trace anomaly proportional to
$R^{(0)ij}R^{(0)}_{ij}-\frac{1}{3}R^{(0)2}$ and corresponds to the local form of the
anomalous part of the effective action associated with the Super Yang-Mills theory in $d=4$
(\cite{Balasubramanian:2000pq},\cite{Liu:1998bu}). Also from \cite{Bautier:2000mz} we know
that this field encodes part of the degrees of freedom contained in the traceless part of
$g^{(4)}$ which, along with $g^{(0)}$, contains all the degrees of freedom of the solutions
for pure gravity in five dimensions. This calculation confirms the previous result obtained
in \cite{Balasubramanian:2000pq} and in \cite{Liu:1998bu} by a different method for the pure
gravitational sector. Our strategy
appears to be more direct than the ones used in the works cited before, however the algebra involved is more complex.\\

The induced four dimensional action we have found here can be considered as a quantum
correction for the Einstein Hilbert action in $d=4$. Mottola and Vaulin \cite{Mottola:2006ew}
have considered a similar idea. They consider these terms as deviations from the classical
stress tensor coming from quantum corrections. We consider to address this problem in a
future work. In another context, this action could be used as an ansatz for the action
proposed in \cite{Kanno:2003vf} to test Kaluza-Klein corrections in the Randall-Sundrum
 two-brane system.

\appendix

\vspace{1in}

{\bf Acknowledgments}

R.A. would like to thank Abdus Salam International Centre for Theoretical Physics (ICTP) for
its support. We also thanks M. Ba\~nados for pointing to us a relevant reference concerning
this work. This work was partially funded by grants FONDECYT 1040202 and DI 06-04 (UNAB) and
(NZ) FONDECYT 1000961.

\providecommand{\href}[2]{#2}\begingroup\raggedright

\endgroup

\end{document}